\documentclass[amsmath,superscriptaddress,amssymb]{revtex4}

\begin{document}

\def\beqa{\begin{eqnarray}}
\def\eeqa{\end{eqnarray}}
\def\beqn{\begin{equation}}
\def\eeqn{\end{equation}}
\def\eqa#1{\begin{eqnarray}\mathnormal{ #1}\end{eqnarray}}
\def\eqn#1{\begin{equation} \mathnormal{#1}\end{equation}}

\def\t{t}
\def\x{x}
\def\xb{{\bf x}}
\def\u{u}
\def\v{v}
\def\k{k}
\def\kb{{\bf k}}
\def\om{\omega}

\def\ph{\phi}
\def\vph{\varphi}
\def\F{F}
\def\a{a}
\def\b{b}
\def\si{\sigma}
\def\ep{\epsilon}

\def\e{e}
\def\T{T}
\def\M{M}
\def\P{P}

\def\J{J}
\def\C{C}
\def\D{D}
\def\Da{\Delta_a}
\def\dDa{{\dot \Delta}_a}
\def\da{\delta_a}
\def\ds{\sigma}
\def\N{N}
\def\al{\alpha}
\def\U{U}
\def\Om{\Omega}
\def\X{X}
\def\S{S}
\def\s{s}
\def\Q{Q}

\title{Conformal symmetry and quantum localization in space-time}
\author{Marc-Thierry Jaekel}
\address{Laboratoire de Physique Th\'eorique de l'Ecole Normale Sup\'{e}rieure, 
CNRS, UPMC, 24 rue Lhomond, F75231 Paris Cedex 05}
\author{Serge Reynaud}
\address{Laboratoire Kastler Brossel, Universit\'{e} Pierre et
Marie Curie,  CNRS, ENS, case 74, Campus Jussieu, F75252 Paris Cedex 05}

\begin{abstract}
The classical procedures which define the relativistic notion of
space-time can be implemented in the framework of Quantum Field Theory. 
Only relying on the conformal symmetries of field propagation, 
time-frequency transfer and localization lead to the definition
of time-frequency references and positions in space-time as quantum observables.
Quantum positions have a non vanishing commutator identifying with spin, both observables
characterizing quantum localization in space-time.
Frame transformations to accelerated frames differ from their classical counterparts.
Conformal symmetry nevertheless allows to extend the  covariance rules underlying the formalism of general relativity under an algebraic form suiting the quantum framework.
\end{abstract}
\maketitle

\section{Introduction}

Different  domains of physics actually use  different representations of the notions of time and space.
The space-time parameter manifold underlying modern physical theories, such as quantum field theories
or general relativity, is alien to the physical observables used by modern metrology 
to build reference systems in space-time.  This situation can  hardly be sustained, 
as it is the source of strong difficulties
when domains lying at the interface between general relativity,
quantum theory and metrology. are being explored with an ever increasing accuracy in 
space and time measurements. 

The different notions of time and space underlying physics were first clearly 
stated by Newton in his {\it Principia} \cite{Newton}~:

"I. Absolute, true, and mathematical time,
of itself, and from its own nature,
flows equably without relation to anything external,
and by another name is called duration.
 Relative, apparent and common time,
is some sensible and external measure
of duration by the means of motion,
which is commonly used instead of true time.

II. Absolute space, in its own nature,
without relation to anything external,
remains always similar and immovable.
Relative space is some movable dimension 
or measure of the absolute spaces,
which our senses determine by its position to bodies, 
and which is commonly taken for immovable space."

The representation of time and space as mathematical parameters, independent of physical observables,
was privileged by Newton and  still lies at the basis of the differential formalism of
modern physical theories. On another hand, when he founded relativity \cite{Einstein05}, Einstein insisted on the necessity to connect space and time to physical observables 
in order to reach a consistent theoretical framework. 
This program is nowadays implemented by  specific physical systems, namely clocks and light signals,
which deliver and disseminate time observables,
and specific procedures, clock synchronization and event localization,
to coordinate events both in time and in space.
The relativistic conception of
space-time and its  constructive procedures lie at the heart of
modern metrology \cite{Guinot77} and coordination systems, 
such as the Global Positioning System (GPS)
\cite{IEEE91,Bahder01}.

The relativistic conception must however face the quantum nature of 
physical observables. The incompatibility of the notions of time and space 
used in the relativistic and quantum formalisms has been early
recognized by Schr\"odinger \cite{Schrodinger}, reviving the fundamental
distinction made by Newton.
As underlined by Schr\"odinger, the formalism of quantum mechanics
preserves  Newton mathematical time while it represents space positions
by operators, a difference of  status which cannot be accepted from the point of view of relativity. 
 Even if space positions may be given 
a representation in terms of quantum operators conjugate to momentum
\cite{Pryce48,NewtonW49,Fleming65},
it is commonly admitted that time cannot be given a similar description as an
operator conjugate to energy \cite{Pauli58,Wightman62}.
Modern Quantum Field Theory manages to restore compatibility, but
at the price of representing both time and space as parameters
and of losing their connection with physical observables. 
This choice is at the origin of  strong conflicts between relativistic and quantum
requirements when attempting to built a consistent theory at their interface 
\cite{Rovelli91,Isham97}.  

In fact, as we discuss here, the relativistic approach
of space-time may be implemented within the standard framework of Quantum Field Theory, 
by applying the latter to a suitable line of reasoning
\cite{JR96a,JR96b,JR98a,JR98b,JR99a,JR04}. Let us first briefly recall the steps
followed by a relativistic construction of space-time.
It begins with a local definition of  time, as a physical observable 
delivered by a clock located at a give place in space \cite{Einstein05}. 
In today's applications, this local time is provided by the most
 accurate clocks available, {\it i.e.} by atomic clocks \cite{IEEE91}.
The local notion of time must then be extended over all space or, equivalently,
the different local times associated with remote clocks must be synchronized. 
For that purpose, two observers share a piece of information which allows them to compare the
indications of their respective local clocks. 
This is accomplished by the exchange of propagating electromagnetic fields. Radio links are used by 
today's most efficient systems for disseminating time references or for 
synchronizing clocks located in satellites (as in the GPS constellation)
or in stations on the Earth surface \cite{IEEE91}. 
A first observer encodes a time reference on an electromagnetic pulse, 
representing the time delivered by his clock.
Comparing the received time reference with the indications of his own clock,
the second observer then proceeds to the identification of the two time
variables,  {\it i.e.} to the synchronization of his clock. To ensure a faithfull comparison,
encoding should be performed using physical quantities which are preserved by field propagation.

Space coordination follows time transfer or synchronization.
An event is completely localized both in space and time
using several transfers of time, at least the same number as the 
space-time dimension. Concrete realizations 
like the GPS \cite{IEEE91,Bahder01} use
a higher number for raising degeneracies between solutions.
An event in space-time is then defined by merging several exchanges of
electromagnetic signals.
Classically, the positions of an event, both in space and time,
is deduced from the different transfered time references.
This is  illustrated in two-dimensional space-time by simple relations 
\beqa
\label{2d_localization}
\t - \x/c = \u^-
\qquad \qquad \t + \x/c = \u^+
\eeqa
The values of light-cone variables ($\u^\pm$) are the quantities preserved by field propagation
and they are sufficient to provide the positions localizing an event in space-time. 
The implementation of these procedures with quantum fields hits upon fundamental limitations 
imposed by quantum theory.  Heisenberg inequalities limit
the possibility to focus field energies at a given value of a light-cone variable.
References used in a time transfer should be represented as non commuting operators defined in 
a quantum algebra.
Transfer and localization should then be performed using quantum observables which are preserved by field propagation. 
Such physical quantities exist, due to the symmetries of field propagation
equations.

In the following, we show how the conformal symmetries underlying the propagation of
quantum fields allow one to realize a relativistic space-time construction 
and to obtain time-frequency references and space-time positions as quantum observables.
We briefly discuss their resulting properties under relativistic transformations
and point at consequences for a potential extension of the formalism founding general relativity.

\section{Quantum time-frequency transfer}

As a first step, we show how observables used in time-frequency transfer or synchronization 
can be implemented in the framework of Quantum Field Theory. 
In order to discuss the quantum properties induced by fields on 
 transfer or synchronization observables, it will be sufficient to use 
a simplified model in terms of a real scalar quantum field in two-dimensional space-time. 
This simplification  amounts to project fields on  the direction of
propagation and hence  to discard all effects related to transverse 
directions or to photon polarizations \cite{JR96a}. 
Such effects, which do not affect time-frequency transfer in an essential way,
 will be taken into account later on, 
when discussing localization in terms of electromagnetic fields in four-dimensional space-time. 

The quantum field used in time-frequency transfer, denoted by $\varphi$, satisfies propagation equations
and can be decomposed over components propagating in opposite directions
\beqa
\label{propagation}
(\partial_\t^2-\partial_\x^2)\vph(\t,\x) = 0 \quad \Leftrightarrow \quad
\vph(\t,\x) = \vph_-(\t-\x) + \vph_+(\t+\x)
\eeqa
Field components themselves decompose over positive and negative frequencies
($\hbar$ is Planck constant divided by $2\pi$, and $\theta$ is Heaviside step
function)
\beqa
\label{spectral_decomposition}
\vph_\si(\u) = \int_{-\infty}^\infty {d\om \over 2\pi} 
\sqrt{\hbar \over 2\om} \theta(\om)  \lbrace  \a_{\om\si}e^{-i\om \u}
+ \a^\dagger_{\om\si} e^{i\om\u}\rbrace
\eeqa
For each propagation direction $\si$, field amplitudes $\a_{\om\si}$ for positive frequency  (respectively $\a_{\om\si}^\dagger$ for negative frequency) act as 
annihilation (respectively creation) operators for photons. Field amplitudes then 
do not commute for equal frequencies 
but satisfy canonical commutation rules ($\delta$ is Kronecker symbol or Dirac distribution, $[A,B]=AB-BA$ denotes the commutator of $A$ and $B$)
\beqa
\label{commutation_rules}
&&[\a_{\om\si}, \a^\dagger_{\om^\prime\si^\prime}] 
= 2\pi \delta_{\si\si^\prime}\delta(\om - \om^\prime), \qquad
[\a_{\om\si}, \a_{\om^\prime\si^\prime}] =  [\a^\dagger_{\om\si}, \a_{\om^\prime\si^\prime}^\dagger] 
=  0
\eeqa
These commutation rules are sufficient
to determine all commutation properties of operators
built on fields, in particular the algebra of observables.

Propagation preserves some quantities which are built on the field and can be obtained from 
 the field energy-momentum tensor ($:\quad:$ denotes a normal product, so that 
the energy-momentum tensor vanishes in vacuum)
\beqa
\label{energy_momentum}
&&\e_\si(\u) = : (\partial_\u\vph_\si(\u))^2 :\nonumber\\
&&\P_\si = \int_{-\infty}^\infty \e_\si(\u) d\u, \qquad 
\J_\si = \int_{-\infty}^\infty \u \e_\si(\u) d\u, \qquad
\C_\si = \int_{-\infty}^\infty \u^2 \e_\si(\u) d\u
\eeqa
The three conserved quantities $\P_\si, \J_\si, \C_\si$ respectively describe, for each propagation direction, the momentum, angular momentum and quadripolar momentum of the quantum field. According to Noether theorem, these
conserved quantities correspond to symmetries of propagation equations 
(\ref{propagation}) and identify with the generators of these symmetries.
 The action on fields of these conserved quantities is deduced from canonical commutation rules
 (\ref{commutation_rules}) 
and  indeed identifies with the action of symmetry generators
\beqa
&&{i\over\hbar}[\P_\si, \vph_\si(\u)] = \partial_\u\vph_\si(\u), \quad 
{i\over\hbar}[\J_\si, \vph_\si(\u)] =  \u \partial_\u\vph_\si(\u), \quad
{i\over\hbar}[\C_\si, \vph_\si(\u)] = \u^2 \partial_\u\vph_\si(\u)\nonumber\\
\eeqa
For each field component, the algebra generated by conserved quantities coincides with
a conformal algebra. In two-dimensional space-time, the conformal symmetry algebra
contains an infinite number of generators (describing arbitrary changes of the light cone variable $\u$).
But only the three momenta written in (\ref{energy_momentum}) will allow for a generalization to 
four dimensions and will be needed for defining localization observables.
These three conserved quantities generate a special conformal algebra
\beqa
\label{conformal_algebra}
&&{i\over\hbar}[\P_\si, \J_\si] = \P_\si, \qquad {i\over\hbar}[\P_\si, \C_\si] = 2 \J_\si, \qquad {i\over\hbar}[\J_\si, \C_\si] = \C_\si
\eeqa
These generators (\ref{energy_momentum}) respectively represent translations,
dilations or Lorentz transformations, and transformations to accelerated frames. 
At this point, it should be noted that the photon number $\mathnormal{\N_\si}$ is preserved by these conformal transformations \cite{JR96a}
\beqa
\label{photon_number}
&&\N_\si = \int_0^\infty {d\om \over 2\pi} \a^\dagger_{\om\si} \a_{\om\si}
\nonumber\\
&&[\P_\si, \N_\si] = [\J_\si, \N_\si] = [\C_\si, \N_\si] = 0
\eeqa
The photon number for each component is defined so that the corresponding energy spectral density
identifies with Planck constant (see eq. (\ref{spectral_decomposition})), so that 
it is a non local expression of the light cone variable decribing the quantum field.
It is invariant under frame transformations corresponding to translations,
Lorentz transformations and accelerations (\ref {energy_momentum}). This means in particular that 
observers which can be related by such frame transformations will give the same physical interpretation
to processes involving photons. Other invariants are also obtained
from the Casimir invariants of the conformal algebra (\ref{conformal_algebra}) ($\cdot$ denotes the symmetrized product)
\beqa
\label{Casimir}
&&{\al_\si}^2 = \C_\si\cdot\P_\si - {\J_\si}^2 
\eeqa
Using the definition of conserved quantities (\ref{energy_momentum}), one sees 
 that  the Casimir invariant $\al_\si^2$
(\ref{Casimir})  measures the field dispersion transversely to the direction of propagation.

In a classical framework, each field component may be seen as propagating while preserving
the value of a light cone variable $u$, which can then be used as
the  reference to be shared by remote observers.
The energy of classical fields can be focussed at a given value of $\u$ with a 
precision which is only limited by technology. 
This allows one to encode a time reference provided by $\u$
and which can be defined, at least in principle, with an arbitrary precision. 
This situation does not hold any more
in the quantum framework, due to Heisenberg inequalities which constrain conjugate 
degrees of freedom, hence field and energy densities. 
The quantum framework nevertheless allows one to define quantum observables  $\Om_\si$ and
 $\U_\si$ representing 
the frequency and the 
position  of a propagating field component and taking the place of the classical frequency $\om$ and the classical light cone variable $\u$ \cite{JR96a}. 
Quantum definitions generalize the classical ones while taking into account the fundamental relations entailed by Planck constant \cite{Einstein05} between energy and frequency, and by relativity \cite{Einstein06} between the energy barycenter (center of inertia) 
and Lorentz boost
\beqa
\label{transfer_observables}
&&\Om_\si = {\P_\si \over \hbar \N_\si}, \qquad \U_\si = \J_\si \cdot {1\over \P_\si} 
\eeqa
The relativistic  quantum  observables defined by (\ref{transfer_observables}) 
provide the required quantum references to be used in  frequency or time transfer.
The usual quantum limits encountered in time and frequency tranfers by means of quantum fields
may then be described in a
universal way by the quantum commutator of the time and frequency observables  
\beqa
\label{transfer_commutator}
[\Om_\si, \U_\si] = -{i\over\N_\si}
\eeqa
The quantum commutator (\ref{transfer_commutator}) 
limiting time and frequency transfers only depends on the photon number,
so that it is identical for different observers which can be related through a conformal frame transformation (\ref{energy_momentum}).
As expected, the quantum references defined by (\ref{transfer_observables}) undergo relatvistic transformations which follow the classical ones in the case of translations and  Lorentz transformations 
\beqa
\label{Poincare_transfer}
&&{i\over\hbar}[\P_\si, \Om_\si] = 0, \qquad {i\over\hbar}[\P_\si, \U_\si] = 1\nonumber\\ 
&&{i\over\hbar}[\J_\si, \Om_\si] = -\Om_\si, \qquad {i\over\hbar}[\J_\si, \U_\si] = \U_\si
\eeqa
The first equations in (\ref{Poincare_transfer}) represent the conjugation relation 
between positions and momenta, 
while the second ones correspond to the Doppler shift and time dilation which characterize Lorentz transformations.
Furthermore, the field propagation equations (\ref{propagation})
 admit a larger group of symmetries than the Weyl group defined by equations 
 (\ref{Poincare_transfer}).
These symmetries include in particular 
transformations between observers in relative acceleration. The corresponding transformations of quantum references are then deduced from their definition (\ref{transfer_observables}) and  the conformal algebra (\ref{conformal_algebra})
\beqa
\label{2d_acceleration_transformation}
&&{i\over\hbar}[\C_\si, \Om_\si] =-2\U_\si\cdot\Om_\si, \qquad
{i\over\hbar}[\C_\si, \U_\si] = {\U_\si}^2 - (\al_\si^2  +{\hbar^2\over4}){1\over\P_\si^2}
\eeqa
The relativistic transformation of the frequency observable corresponds to a position dependent shift of frequency, thus providing a quantum generalization of the classical redshift undergone by frequencies when seen by an accelerated observer.
The first relation in (\ref{2d_acceleration_transformation}) then 
realizes a quantum version of Einstein effect on clocks in an accelerated frame
\cite{Einstein07}. Meanwhile, the relativistic transformation of the position observable
 no longer identifies with its classical analog but involves
an additional term which may be seen as reflecting the fundamental constraints entailed on field dispersions by
Heisenberg inequalities. Using equations (\ref{transfer_observables})), 
the Casimir invariants $\al_\si^2$
(\ref{Casimir}) allow one to rewrite the special conformal generators $\C_\si$
in terms of positions and momenta 
\beqa
\label{accelation_generator}
\C_\si = \U_\si \P_\si \U_\si + (\al_\si^2 +{\hbar^2\over4}){1\over\P_\si}
\eeqa
Last relations in (\ref{2d_acceleration_transformation}) then follow from conjugation (\ref{Poincare_transfer}) and relation (\ref{accelation_generator}).
It can be shown that the term ${\hbar^2 /4}$ corresponds to a minimum of 
field dispersions, the latter being attained for single photon states \cite{JR96a}.

The scheme described in the previous section may be applied using quantum fields. 
Combining time tranfers in different directions
allows one to define localization observables for positions in space and time in terms of quantum fields.
In a two-dimensional space-time, relations (\ref{2d_localization}) may be used and
relativistic transformations of localization observables simply follow from those of transfer observables
(\ref{Poincare_transfer}) and (\ref{2d_acceleration_transformation}), that is from the whole conformal algebra generated by both propagation directions. In fact, the properties of quantum localization observables quite generally follow from the conformal algebra, as discussed in next section.

\section{Quantum localization}

In order to provide a more realistic description of
quantum localization observables in four-dimensional space-time,
 we now discuss their implementation  using electromagnetic
fields, thus taking into account polarization properties  \cite{JR04}.

Electromagnetic  fields satisfy Maxwell equations and the solutions
may be decomposed as plane waves, with wave vectors $\k$ lying on the light-cone ($\k^2=0$)
and corresponding amplitudes ($\a_{\k\si}, \a^\dagger_{\k\si}$) themselves
decomposing over  two helicity components (${\si=\pm}$) which describe photon polarizations
\cite{ItzyksonZ85}.
 To circumvent gauge ambiguitites, electromagnetic fields will be described
 by their real antisymmetric tensor 
\beqa
\label{Maxwell_fields}
&&\F_{\mu\nu}(x) = \int {d^4\k \over(2\pi)^3}\delta(\k^2) \theta(\k_0)\Sigma_{\si} \lbrace \ep_{\mu\nu}^\si(\k) \a_{\k\si}e^{-i\k\x} + \ep_{\mu\nu}^{\si}(\k)^*\a^\dagger_{\k\si}  e^{i\k\x}\rbrace
\eeqa
$\ep_{\mu\nu}^\si(\k)$ denotes the expression taken by the polarization tensor of mode  $(\k,\si)$, as determined by Maxwell equations, and $*$ denotes complex conjugation.

The algebra of quantum fields is obtained from the canonical commutation rules satisfied
by photon creation ($\a^\dagger_{\k\si}$) and annihilation  ($\a_{\k\si}$) operators
\beqa
\label{field_canonical_commutation_rules}
&&[ \a_{\k\si}, \a^\dagger_{\k^\prime\si^\prime} ]= (2\pi)^3 \delta_{\si \si^\prime}(2\k_0)\delta^3(\k-\k^\prime), \quad
[ \a_{\k\si}, \a_{\k^\prime\si^\prime} ]=[ \a^\dagger_{\k\si}, \a^\dagger_{\k^\prime\si^\prime} ]= 0 
\eeqa
($(2\k_0)\delta^3(\k-\k^\prime)$ stands for the invariant Dirac distribution on the light cone).
 
Due to the conformal invariance of Maxwell equations \cite{Bateman09,Cunningham09}, the electromagnetic field energy-momentum tensor $\T_{\mu\nu}$ provides quantities $\Da$   which are preserved by propagation  and which correspond to generators of conformal symmetries \cite{BinegarFH83} (indices are lowered and raised using Minkowski metric  $\eta_{\mu\nu}$; $\Da$ is independent of $\t$)
\beqa
\label{conserved_quantities}
&&\T_{\mu\nu}(\x) = :\F_{\lambda\mu}(\x){\F_\nu}^\lambda(\x) + {1\over4}\eta_{\mu\nu}\F_{\lambda\rho}(\x)\F^{\lambda\rho}(\x):, 
\qquad \eta_{\mu\nu} = {\rm diag}(1, -1, -1, -1)\nonumber\\
&&\Da = \int \delta(\x^0-t)  \T_{0\mu}(\x) \da^\mu (\x) \, d^4 \x 
\eeqa

The set of space-time functions $\da^\mu$ which determine the conserved quantities
is fixed by the symmetries of the light cone ($\eta_{\mu\nu}(\x^\mu+\da^\mu(\x))(\x^\nu+\da^\nu(\x)) = \eta_{\mu\nu}\x^\mu\x^\nu=0$) and a basis is given by
the following 15 expressions ($a=\P_{\mu},\J_{\mu\nu},\D,\C_\mu$, for  $\mu,\nu = 0,1,2,3$)
\beqa
\label{conformal_generators_expressions}
&&\delta_{\P_\nu }^\mu (\x)=\delta_\nu^\mu\nonumber\\
&&\delta_{\J_{\nu \rho }}^\mu (\x)=\delta_\nu^\mu \x_\rho -\delta_\rho^\mu \x_\nu, \qquad
\delta_\D^\mu (\x)=\x^\mu\nonumber\\
&&\delta_{\C_\nu }^\mu (\x)=2\x_\nu \x^\mu -\delta_\nu ^\mu \x^2
\eeqa
From canonical field commutation relations (\ref{field_canonical_commutation_rules}), 
the conserved quantities $\Da$ defined by relations (\ref{conserved_quantities}) generate conformal transformations of the fields. In fact, Maxwell fields defined by (\ref{Maxwell_fields}) correspond
to particular representations (with helicity $\pm 1$) of the conformal algebra (\ref{conformal_generators_expressions}) on the light cone \cite{Bracken71,Bracken81}
\beqa
\label{field_conformal_transformations}
&& {i\over\hbar}[\Da, \F_{\mu\nu}(\x)] = (\da \F)_{\mu\nu} (\x), \qquad 
\da = \da^\mu(\x)\partial_\mu + \ds_a, \qquad  \partial_\mu = {\partial\over\partial\x^\mu}\nonumber\\
&&{i\over\hbar}[\Da, \Delta_b] = \Delta_{[a,b]}, \quad 
\delta _{[a,b]} = \lbrace\delta _a^\nu(\x) \partial _\nu \delta_b^\mu(\x) 
-\delta_b^\nu(\x) \partial _\nu \delta_a^\mu(\x)\rbrace\partial_\mu + [\ds_a, \ds_b]\nonumber\\
\eeqa
$\ds_a$, which describes the polarization dependent part of the conformal generator,
 is a matrix acting on field indices which is determined by Maxwell equations and 
its symmetries (\ref{conformal_generators_expressions}). Relations (\ref{field_conformal_transformations}) allow one to identify the conserved quantities $\Da$
with the corresponding symmetry generators $\a$.

As in the two-dimensional case, 
conformal symmetries describe changes of reference frame. 
In the four-dimensional case,
relativistic transformations include
translations $\mathnormal{\P_\mu}$ and Lorentz transformations $\mathnormal{\J_{\mu \nu }}$,
corresponding to momentum and angular momentum conservations, and
satisfying a Poincar\'e algebra 
\beqa
\label{Poincare_generators}
&&[\P_\mu ,\P_\nu] = 0, \qquad {i\over\hbar}[ \J_{\mu \nu }, \P_\rho ] =
 \eta _{\mu \rho }\P_\nu  -\eta _{\nu \rho }\P_\mu\nonumber\\
&&{i\over\hbar}[ \J_{\mu \nu }, \J_{\rho \sigma }] = 
 \eta _{\mu \rho } \J_{\nu \sigma } + \eta _{\nu \sigma }\J_{\mu \rho } 
-\eta _{\nu \rho } \J_{\mu \sigma} - \eta _{\mu \sigma }\J_{\nu \rho }  
\eeqa
But they also include dilatations generated by  $\mathnormal{\D}$
\beqa
\label{dilation_generator}
{i\over\hbar}[\D, \P_\mu] = - \P_\mu   \qquad \qquad 
[\D, \J_{\mu \nu }] = 0
\eeqa
and  transformations to uniformly accelerated frames $\mathnormal{C_\mu}$
\beqa
\label{acceleration_generator}
&&{i\over\hbar}[\C_\mu , \P_\nu] = 2\left(\J_{\mu \nu } -\eta _{\mu \nu }\D\right) \nonumber\\
&&{i\over\hbar}[\J_{\mu \nu }, \C_\rho] = \eta _{\mu \rho }\C_\nu 
 -\eta _{\nu \rho }\C_\mu, \qquad {i\over\hbar}[\D, \C_\mu] =  \C_\mu\nonumber\\
&&[\C_\mu , \C_\nu] = 0
\eeqa
The photon number provides a  first quantum observable built on Maxwell fields 
 and remaining invariant under conformal transformations (\ref{field_conformal_transformations}).
 It may be written as a quadratic form of electromagnetic fields
(\ref{Maxwell_fields}) and is non local in the space-time parameter $\x$. This invariance ensures that 
observers which are related through  a conformal transformation will have the same interpretation
of processes involving photons.

Classically, positions in four-dimensional space-time are defined 
from the intersection of at least four light cones (for instance emitted by beacons 
located on satellites of the GPS constellation). Equivalently, classical positions are  
defined from the intersection of at least four light rays with different propagation
directions \cite{JR96b}. This generalizes the two-dimensional case (\ref{2d_localization}) and
 provides explicit algebraic expressions relating the position components with 
the total momentum $\P_\mu$, angular momentum $\J_{\mu\nu}$ and dilation generator $\D$ of the field configuration \cite{JR96b}.  The same definition may be applied in the quantum case by considering  
quantum fields merging on the localization event and possessing momenta components corresponding to
different propagation directions. This means  that the quantum fields used for localization 
correspond to
a total squared  mass ($\P^2 =\M^2$) which does not vanish. Generalizing the relations satisfied by classical positions leads to the following definition for quantum positions in space-time 
\beqa
\label{position_observables}
\X_\mu = {1\over\P^2}\cdot\left(\P^\lambda \cdot \J_{\lambda \mu} + \P_\mu \cdot \D\right)
\eeqa
In contrast to the two-dimensional case, the position observables defined by equations
(\ref{position_observables}) do not exhaust all the information contained in the localization fields.
Indeed, spin components may further be defined either under the form of a vector $\S_\mu$ (Pauli-Lubanski vector \cite{ItzyksonZ85}) or of an antisymmetric tensor $\S_{\mu\nu}$
($\epsilon_{\mu\nu\lambda\rho}$ is the completely antisymetric tensor with $\epsilon_{0123}=1$)
\beqa
\label{spin_observables}
\S_\mu = - {1\over 2} \epsilon_{\mu\nu\lambda\rho} \P^\nu \J^{\lambda\rho},
 \qquad \S_{\mu\nu} = \epsilon_{\mu\nu\lambda\rho} {\P^\lambda \over \P^2} \S^\rho
\eeqa
Position and spin observables characterize space-time localization.
In particular, Weyl generators  are recovered from these observables
and take their usual form
\beqa
\J_{\mu \nu} = \P_\mu \cdot \X_\nu - \P_\nu \cdot \X_\mu 
+ \S_{\mu \nu}, \qquad \D = \P^\mu \cdot \X_\mu
\eeqa
As was the case for time-frequency observables (\ref{Poincare_transfer}), quantum positions and momenta correspond to conjugate observables 
\beqa
\label{conjugate_observables}
{i\over\hbar}[\P_\mu , \X_\nu] = \eta_{\mu \nu}
\eeqa
Conjugation relations (\ref{conjugate_observables}) directly follow from the conformal algebra (\ref{Poincare_generators},\ref{dilation_generator}) and the definition of quantum positions
(\ref{position_observables}). In particular, their time component is a direct extension of
the time-frequency conjugation relation (\ref{Poincare_transfer}). These relations then sustain the existence of a time operator conjugate to energy, contrarily to a common opinion \cite{Pauli58,Wightman62}.
This paradox disappears when considering that the position observables defined by equations 
(\ref{position_observables}) correspond to hermitian (or symmetric operators) which are not self-adjoint. This property is to be connected with the important part of the Hilbert space which 
is excluded by definition, in particular all field states with a vanishing total mass.  
Indeed, non self-adjoint operators not only provide quite acceptable representations of 
physical observables \cite{BuschGL95}, but even appear to be unavoidable when dealing with localization 
in a non-commutative space-time \cite{Kempf00}. 
As another characteristic property of quantum positions, one notes that the latter possess
a non vanishing commutator which is related to the spin observable (\ref{spin_observables})
\beqa
\label{position_commutator}
{i\over\hbar}[ \X_\mu , \X_\nu ] =  -{\S_{\mu \nu}\over\P^2}
\eeqa
As can be seen on the simple example  of two-photon states built with electromagnetic fields
(\ref{Maxwell_fields}) \cite{JR99a}, both 
 the spatial dispersion of quantum fields and the photon helicities contribute
 to the position commutator (\ref{position_commutator}).
This commutator shows that quantum localization in space-time is affected by size effects,
characterized by spin in a mass dependent way. This 
points at an intimate relation between position and spin observables and at a scale dependence of the resulting space-time non-commutativity.

Relativistic transformations between observers being determined by conformal symmetries,
the definitions of quantum positions (\ref{position_observables}) and spin (\ref{spin_observables})
result in classical transformations  under Lorentz transformations  (\ref{Poincare_generators})
and dilatation (\ref{dilation_generator})
\beqa
{i\over\hbar}[\J_{\mu \nu} , \X_\rho] = \eta_{\mu \rho} \X_\nu 
 - \eta_{\nu \rho} \X_\mu, \qquad  {i\over\hbar}[\D , \X_\mu ] = \X_\mu\nonumber\\
{i\over\hbar}[\J_{\mu \nu} , \S_\rho] = \eta_{\mu \rho} \S_\nu 
 - \eta_{\nu \rho} \S_\mu, \qquad  {i\over\hbar}[\D , \S_\mu ] = -\S_\mu
\eeqa
However, transformations between observers in relative acceleration exhibit specific behaviors of quantum
localization observables. First, quantum positions are related to the
transformation of the mass observable $\M$ under an acceleration $\a^\mu$ (the mass observable $\M$ being Lorentz invariant) 
\beqa
\label{mass_transformation}
&&\Delta = \frac{\a^\mu}2 \C_\mu\nonumber\\
&&{i\over\hbar}[\Delta , \M ] = - \Phi \cdot \M, \qquad \Phi = \a^\mu \X_\mu
\eeqa
As in the classical case, the accelerated mass  (\ref{mass_transformation}) undergoes a shift proportional to position similar to the redshift undergone by frequency references (\ref{2d_acceleration_transformation}). The mass transformation (\ref{mass_transformation}) may also be
used as an equivalent definition of quantum positions (\ref{position_observables}). 

As a consequence of the conformal algebra (\ref{acceleration_generator}) and of the definitions of positions and spin observables (\ref{position_observables},\ref{spin_observables}), transformations between frames in relative acceleration mix the different quantum localization observables
\beqa
\label{acceleration_transformations}
\Delta &=& {\a^\mu \over2}\left(2 (\P\cdot\X)\cdot\X_\mu - \X^2\cdot\P_\mu +2\X^\nu\S_{\nu\mu}-{\S^2\over(\P^2)^2}\P_\mu + 2\Q_\mu\right)
\nonumber\\
{i\over\hbar}[\Delta, P_\mu] &=&  \left(\a^\nu\X_\mu - \a_\mu\X^\nu -\a\X\delta^\nu_\mu\right)\cdot \P_\nu -\a^\nu\S_{\mu\nu}\nonumber\\ 
{i\over\hbar}[\Delta, \X_\mu] &=& \a^\nu \left( \X_\nu  \cdot \X_\mu- {\eta_{\mu\nu} \over 2}\X^2\right) 
-{\a^\nu\over\P^2}\left({\S_\nu \cdot \S_\mu\over\P^2} -{\eta_{\mu\nu}\over2}{\S^2\over\P^2}\right)\nonumber\\
&& +\left(\a_\mu\Q^\nu - \a^\nu\Q_\mu -\a\Q\delta^\nu_\mu\right)\cdot {\P_\nu\over\P^2}\nonumber\\
{i\over\hbar}[ \Delta, \S_\mu] &=&   \left(\a^\nu\X_\mu - \a_\mu\X^\nu -\a\X\delta^\nu_\mu\right)\cdot \S_\nu -\epsilon_{\mu\nu\rho\sigma}\a^\nu\P^\rho\Q^\sigma
\eeqa
Besides positions and spin, quadrupole momenta $\Q_\mu$ appear when
rewriting the acceleration  generator $\Delta$ in terms of localization observables \cite{JR99a}.
Comparing expressions (\ref{acceleration_transformations})  with their classical analogs shows that
 the transformation of momentum
contains an additional contribution related to spin, a consequence of intrinsic 
quantum dispersions (\ref{position_commutator}). The intimate relation between positions and spin also 
manifests itself in their mixed transformations. Similarly to the
transformations of time references (\ref{2d_acceleration_transformation}),
transformations of positions involve corrections with respect to classical expressions
 which depend on spin, quadrupole momenta and  momentum \cite{JR99a}.  

At the classical level, relativistic transformations between observers in relative acceleration are 
usually treated in the framework of  differential geometry.
The non-commutative character of quantum positions (\ref{position_commutator}) signals a failure,
at the quantum level, of the  classical properties relating fields and space-time coordinates.
 In particular, the non vanishing
commutator of quantum positions precludes the use of 
classical covariance rules for representing frame transformations. This can also be seen in the occurence of spin and quadrupole dependent terms in the transformation of quantum positions (\ref{acceleration_transformations}).
However, equations (\ref{acceleration_transformations}) show that, 
besides position independent corrections, transformations under acceleration of spin and momentum  are given by the same
 position dependent linear operator, which takes the same form
as its classical analog \cite{JR96b}
\beqa
\label{acceleration_covariance}
\delta^\mu_\Delta(\x) &=& \a^\nu(\x_\nu \x^\mu -{\delta_\nu ^\mu\over2}  \x^2)\nonumber\\
{i\over\hbar}[\Delta, P_\mu] &=& -\partial_\mu\delta^\nu_\Delta(\X)\cdot \P_\nu -\a^\nu\S_{\mu\nu}\nonumber\\
{i\over\hbar}[ \Delta, \S_\mu] &=&  -\partial_\mu\delta^\nu_\Delta(\X) \cdot \S_\nu -\epsilon_{\mu\nu\rho\sigma}\a^\nu\P^\rho\Q^\sigma
\eeqa
Similarly, besides position independent corrections, positions transform classically, provided a symmetrized product is used, and the spin-quadrupole independent
part of positions shift has for differential the previous linear operator
\beqa
\label{acceleration_covariance2}
\delta^\mu_\Delta(\X) &=& \a^\nu \left( \X_\nu  \cdot \X^\mu- {\delta^\mu_\nu \over 2} \X^2\right)\nonumber\\
{i\over\hbar}[\Delta, \X^\mu] &=& \delta^\mu_\Delta(\X) - \delta^\mu_\Delta({\S\over\P^2})
-\partial_\nu\delta^\mu_{\Delta}(\Q)  {\P^\nu\over\P^2}
\eeqa
A trace of classical covariance properties remains  at the quantum level,
suggesting a possible generalization. Indeed, the algebraic framework of 
quantum theory  bears constitutive rules which allow extensions of the covariance rules.
As a  consequence of Jacobi identities applied to the conformal 
algebra and of the invariance of the commutator between momenta and positions,
 the following identities are satisfied
\beqa
\label{quantum_covariance}
&&\left[\P_\mu , \left[ \Delta, \X_\nu \right] \right] -
\left[ \X_\nu , \left[ \Delta, \P_\mu \right] \right] = \left[\Delta , \left[ \P_\mu, \X_\nu \right] \right] = 0
\eeqa
The first commutator in the first equation in (\ref{quantum_covariance}) isolates the position dependent part in the position shift and takes its differential while the second commutator isolates the momentum dependent part in the  momentum shift. The algebraic relation (\ref{quantum_covariance}) 
may then be seen as an algebraic expression of the relation between equations (\ref{acceleration_covariance}) and (\ref{acceleration_covariance2}) \cite{JR96b}.

Commutators between momenta and position shifts appearing in (\ref{quantum_covariance}) have a direct physical interpretation, as they correspond to the shifts undergone by clock rates under changes of
reference frames. Equations (\ref{quantum_covariance}) extend at the quantum level the classical relations between clock rates and frequency shifts according to Einstein effect.
Symmetrizing the expressions of these commutators in their indices provides a 
symmetric tensor which, in the particular case of accelerated frames,
 extends at the quantum level the corresponding conformal factor (\ref{mass_transformation}) 
\beqa
\label{quantum_metric}
\left[ \P_\mu , \left[ \Delta, \X_\nu \right] \right] +
\left[ \P_\nu , \left[ \Delta, \X_\mu \right] \right] =
\left[ \X_\mu , \left[ \Delta, \P_\nu \right] \right] +
\left[ \X_\nu , \left[ \Delta, \P_\mu \right] \right] =
-2\hbar^2 \Phi \eta_{\mu \nu}
\eeqa
Relation (\ref{quantum_metric}) may be seen as a quantum extension of the  classical notion of metric \cite{JR96b}.

\section{Conclusion}

   We have shown that the classical procedures which define the relativistic notion of
space-time and which are used in practical realizations of reference frames
can be implemented in the framework of Quantum Field Theory. 
When expressed by means of quantum fields,
time-frequency transfer, or synchronization, and localization lead to the definition
of references and positions in space-time as quantum observables. This success meets the necessity
imposed by relativity and quantum theory to represent space-time positioning with physical observables,
hence with quantum operators. 

The basic and sufficient structure required for defining space-time positioning 
appears to be provided by symmetries of field propagation and more precisely conformal symmetries. 
Characteristic properties of relativity, such as Einstein effect, follow from the conformal algebra
\cite{JR96a,JR98b}.
As a bonus with respect to the classical situation, 
symmetry generators identify with quantities preserved by field propagation
so that both relativistic transformations and quantum operators are expressed within a single algebra.
If the use of electromagnetic fields imposes itself on practical grounds, gravitational fields would lead to equivalent space-time positions, showing that 
universality of space-time is preserved at the quantum level. 
The major difference with classical space-time
 appears in the non-commutativity of quantum positions, characterized by spin. 
The latter also enter the transformations of positions under frame transformations
between relatively accelerated observers \cite{JR99a}. Considering spin as a fundamental element
of a quantum extension of space-time opens new perspectives, for instance
in view of applying quantum deformations of space-time symmetries \cite{PodlesW90}  
or for reconciling quantum fields with a mechanical description of elementary particles \cite{JR99b,JR00,JR01}.

Although non-commutativity precludes the ordinary use of differential geometry for quantum positions,
the frame transformations of quantum observables keep the trace of  
the covariance rules which play a crucial role in the formalism of general relativity.
Conformal invariance moreover allows one to
extend, for uniform accelerations, the covariance rules under a purely algebraic form suiting the quantum formalism \cite{JR96b,JR04}. 
This remarkable property points at an alternative quantum extension of the 
metric fields founding general relativity as a gravitation theory.

\end{document}